\documentclass[twocolumn]{aastex7}

\usepackage{amsmath}
\newcommand\eleven{$11.3~\micron$}
\newcommand\seven{$7.7~\micron$}
\newcommand\six{$6.2~\micron$}
\newcommand\three{$3.3~\micron$}
\newcommand\ElevenSeven{$11.3/7.7~\micron$}
\newcommand\ThreeSeven{$3.3/7.7~\micron$}
\newcommand\ThreeEleven{$3.3/11.3~\micron$}
\newcommand\kms{km~s$^{-1}$}
\newcommand\ha{H$\alpha$}
\newcommand\LTIR{$L_{\rm TIR}$}
\newcommand\LPAH{$L_{\rm PAH}$}
\newcommand\qpah{$q_{\rm PAH}$}
\newcommand\Msun{M$_{\odot}$}
\newcommand\Msunperyear{M$_{\odot}$~yr$^{-1}$}
\newcommand\LHands{L.~Hands et al.~(in prep)}

\shorttitle{PAH Evolution in the NGC~253 Wind}
\shortauthors{S.~A.~Cronin}

\received{November 2026}
\revised{XXX}
\accepted{XXX}

\submitjournal{ApJ Letters}

\graphicspath{{./}{figures/}}

\begin{document}

\title{JWST Observations of Starbursts: Polycyclic Aromatic Hydrocarbons in the NGC~253 Superwind}

\author[0000-0002-9511-1330]{Serena A. Cronin}
\affiliation{Department of Astronomy, University of Maryland, College Park, MD 20742, USA}
\email[show]{cronin@umd.edu}

\author[0000-0002-5480-5686]{Alberto D. Bolatto}
\affiliation{Department of Astronomy, University of Maryland, College Park, MD 20742, USA}
\affiliation{Joint Space-Science Institute, University of Maryland, College Park, MD 20742, USA}
\email{bolatto@umd.edu}

\author[0000-0003-2508-2586]{Rebecca C. Levy}
\affiliation{Space Telescope Science Institute, 3700 San Martin Drive, Baltimore, MD 21218, USA}
\email{rlevy.astro@gmail.com}

\author[0000-0001-5340-6774]{Karl D. Gordon}
\affiliation{Space Telescope Science Institute, 3700 San Martin Drive, Baltimore, MD 21218, USA}
\email{kgordon@stsci.edu}

\author[0000-0003-1356-1096]{Elizabeth Tarantino}
\affiliation{Space Telescope Science Institute, 3700 San Martin Drive, Baltimore, MD 21218, USA}
\email{etarantino@stsci.edu}

\author[0000-0003-4850-9589]{Martha L. Boyer}
\affiliation{Space Telescope Science Institute, 3700 San Martin Drive, Baltimore, MD 21218, USA}
\email{mboyer@stsci.edu}

\author[0009-0005-0750-2956]{Lindsey Hands}
\affiliation{Department of Astronomy \& Astrophysics, University of California, San Diego, 9500 Gilman Drive, La Jolla, CA 92093, USA}
\email{lhands@ucsd.edu}

\author[0009-0001-5949-1524]{Hannah B. Koziol}
\affiliation{Department of Astronomy \& Astrophysics, University of California, San Diego, 9500 Gilman Drive, La Jolla, CA 92093, USA}
\email{hkoziol@ucsd.edu}

\author[0000-0002-2545-1700]{Adam~K.~Leroy}
\affil{Department of Astronomy, The Ohio State University, 140 W. 18th Ave., Columbus, OH 43210, USA}
\affil{Center for Cosmology and AstroParticle Physics, The Ohio State University, 191 W. Woodruff Ave., Columbus, OH 43210, USA}
\email{leroy.42@osu.edu}

\author[0009-0001-6844-9758]{Patricia A. Arens}
\affiliation{Centre for Astrophysics and Supercomputing, Swinburne University of Technology, Hawthorn, VIC 3122, Australia}
\email{parens@swin.edu.au}

\author[0000-0002-3952-8588]{Leindert A. Boogaard} 
\affiliation{Leiden Observatory, Leiden University, PO Box 9513, NL-2300 RA Leiden, The Netherlands}
\email{boogaard@strw.leidenuniv.nl}

\author[0000-0002-5782-9093]{Daniel~A.~Dale}
\affiliation{Department of Physics and Astronomy, University of Wyoming, Laramie, WY 82071, USA}
\email{ddale@uwyo.edu}

\author[0000-0002-6460-3682]{Fergus R. Donnan}
\affiliation{Department of Astronomy \& Astrophysics, University of California, San Diego, 9500 Gilman Drive, La Jolla, CA 92093, USA}
\email{fdonnan@ucsd.edu}

\author[0009-0001-6065-0414]{Grant P. Donnelly}
\affiliation{Department of Physics \& Astronomy and Ritter Astrophysical Research Center, University of Toledo, Toledo, OH 43606, USA}
\email{Grant.Donnelly@rockets.utoledo.edu}

\author[0000-0001-6527-6954]{Kimberly L.~Emig}
\affil{National Radio Astronomy Observatory, 520 Edgemont Road, Charlottesville, VA 22903, USA}
\email{kemig@nrao.edu}

\author[0000-0003-0645-5260]{Deanne B. Fisher}
\affil{Centre for Astrophysics and Supercomputing, Swinburne University of Technology, Hawthorn, VIC 3122, Australia}
\email{dfisher@swin.edu.au}

\author[0000-0002-2775-0595]{Rodrigo Herrera-Camus}
\affil{Departamento de Astronomía, Universidad de Concepción, Barrio Universitario, Concepción, Chile}
\affil{Millennium Nucleus for Galaxies (MINGAL), Concepción, Chile}
\email{rodrherrera@udec.cl}

\author[orcid=0000-0002-0560-3172]{Ralf S.\ Klessen}
\affiliation{Universit\"{a}t Heidelberg, Zentrum f\"{u}r Astronomie, Institut f\"{u}r Theoretische Astrophysik, Albert-Ueberle-Str.\ 2, 69120 Heidelberg, Germany}
\affiliation{Universit\"{a}t Heidelberg, Interdisziplin\"{a}res Zentrum f\"{u}r Wissenschaftliches Rechnen, Im Neuenheimer Feld 225, 69120 Heidelberg, Germany}
\email{klessen@uni-heidelberg.de}

\author[0000-0001-8490-6632]{Thomas S.-Y. Lai}
\affil{IPAC, California Institute of Technology, 1200 E. California Blvd., Pasadena, CA 91125, USA}
\email{ThomasLai.astro@gmail.com}

\author[0000-0003-4023-8657, sname='Lenkić']{Laura Lenkić}
\affil{IPAC, California Institute of Technology, 1200 E. California Blvd., Pasadena, CA 91125, USA}
\email{llenkic@ipac.caltech.edu}

\author[0000-0002-1254-4174]{Ashley E. Lieber}
\affiliation{Department of Physics and Astronomy, University of Kansas, 1251 Wescoe Hall Drive, Lawrence, KS 66045, USA}
\email{ashleylieber@ku.edu}

\author[0000-0002-1790-3148]{Laura A. Lopez}
\affil{Department of Astronomy, The Ohio State University, 140 W. 18th Ave., Columbus, OH 43210, USA}
\affil{Center for Cosmology and AstroParticle Physics, The Ohio State University, 191 W. Woodruff Ave., Columbus, OH 43210, USA}
\email{lopez.513@osu.edu}

\author[0000-0002-2644-0077]{Sebastian Lopez} 
\affil{School of Earth and Space Exploration, Arizona State University, P.O. Box 876004, Tempe, AZ 85287, USA}
\email{slope225@asu.edu}

\author[0000-0001-9436-9471]{David S. Meier}
\affiliation{New Mexico Institute of Mining and Technology, 801 Leroy Place, Socorro, NM, 87801, USA}
\email{david.meier@nmt.edu}

\author[0000-0001-8782-1992]{Elisabeth A.C. Mills}
\affiliation{Department of Physics and Astronomy, University of Kansas, 1251 Wescoe Hall Drive, Lawrence, KS 66045, USA}
\email{eacmills@ku.edu}

\author[0000-0001-8224-1956]{Juergen Ott}
\affiliation{National Radio Astronomy Observatory, P.O. Box O, 1003 Lopezville Road, Socorro, NM 87801, USA}
\email{jott@nrao.edu}

\author[0000-0003-2721-487X]{Debosmita~Pathak}
\altaffiliation{NSF Graduate Research Fellow}
\affil{Department of Astronomy, The Ohio State University, 140 W. 18th Ave., Columbus, OH 43210, USA}
\affil{Center for Cosmology and AstroParticle Physics, The Ohio State University, 191 W. Woodruff Ave., Columbus, OH 43210, USA}
\email[]{pathak.89@osu.edu}

\author[orcid=0000-0002-4735-8494]{Sophie L. Robbins}
\affiliation{Department of Astronomy, University of Maryland, College Park, MD 20742, USA}
\email{srobbin7@umd.edu}  

\author[0000-0001-6325-9317]{Helena M. Richie}
\affiliation{Center for Computational Astrophysics, Flatiron Institute, 162 5th Avenue, New York, NY 10010, USA}
\email{helenarichie@gmail.com}

\author[0000-0002-4378-8534]{Karin M. Sandstrom}
\affiliation{Department of Astronomy \& Astrophysics, University of California, San Diego, 9500 Gilman Drive, La Jolla, CA 92093, USA}
\email{kmsandstrom@ucsd.edu}

\author[0009-0005-8382-0614]{Kaitlyn E. Sheriff}
\affiliation{Department of Physics and Astronomy, University of Kansas, 1251 Wescoe Hall Drive, Lawrence, KS 66045, USA}
\email[]{kaitlyn.sheriff@ku.edu}

\author[0000-0003-1545-5078]{J.D.T. Smith}
\affiliation{Department of Physics \& Astronomy and Ritter Astrophysical Research Center, University of Toledo, Toledo, OH 43606, USA}
\email{JD.Smith@utoledo.edu}

\author[0000-0003-4209-1599]{Yu-Hsuan Teng}
\affiliation{Department of Astronomy, University of Maryland, College Park, MD 20742, USA}
\email{yhteng@umd.edu}

\author[0000-0002-3158-6820]{Sylvain Veilleux}
\affiliation{Department of Astronomy, University of Maryland, College Park, MD 20742, USA}
\affiliation{Joint Space-Science Institute, University of Maryland, College Park, MD 20742, USA}
\email{veilleux@umd.edu}

\author[0000-0002-5877-379X]{Vicente Villanueva}
\affiliation{Instituto de Estudios Astrofísicos, Facultad de Ingeniería y Ciencias, Universidad Diego Portales, Av. Ejército Libertador 441, 8370191 Santiago, Chile}
\affil{Millennium Nucleus for Galaxies (MINGAL), Concepción, Chile}
\email{vicente.avl365@gmail.com}

\author[0000-0003-4793-7880]{Fabian Walter}
\affiliation{Max Planck Institut f\"ur Astronomie, K\"onigstuhl 17, D-69117 Heidelberg, Germany}
\affiliation{California Institute of Technology, Pasadena, CA 91125, USA}
\email{walter@mpia.de}

\author[0000-0001-5434-5942]{Paul P.~van der Werf}
\affiliation{Leiden Observatory, Leiden University, PO Box 9513, NL-2300 RA Leiden, The Netherlands}
\email{pvdwerf@strw.leidenuniv.nl}

\begin{abstract}

We present the first spatially resolved study of polycyclic aromatic hydrocarbon (PAH) evolution in the galactic superwind of the nearby ($d \sim 3.5$~Mpc) galaxy NGC~253. Using JWST MIRI and NIRCam, we image $\sim \pm2$~kpc along the galaxy minor axis, capturing both sides of the starburst-driven wind at $\sim0\farcs05{-}0\farcs375$ ($\sim0.85{-}6.4$~pc) resolution and providing one of the most detailed views of the wind to date. We target PAH emission at \three, \seven, and \eleven, revealing PAH-bright filamentary streamers along the limb-brightened edges of the outflow cones out to at least a deprojected distance $r \sim 0.5$~kpc, beyond which they likely become confused with the spiral arms. In the southwest (SW) and northeast (NE) streamers, PAH intensity declines faster than $r^{-2}$, suggesting either PAH destruction or a decrease in gas column density or dust-to-gas ratio rather than a response to declining illumination alone. Band ratios trace changes in PAH properties, with the \ElevenSeven\ ratio increasing with distance in the SW and northwest (NW) streamers, and \ThreeSeven\ and \ThreeEleven\ decreasing in the SW and NE streamers, indicating larger, more neutral PAHs farther from the starburst. Given the thermal and dynamical properties of the gas in the SW streamer, cool gas is expected to survive acceleration in the outflow and enable PAH shielding. Compared with those of the prototypical starburst M82, the NGC~253 band ratios indicate faster evolution in PAH size. Thus, the NGC~253 wind may represent an earlier stage of outflow evolution in which small PAHs are destroyed near the starburst within the first few Myr.

\end{abstract}



\section{Introduction} \label{sec:intro}

Galactic winds, or outflows, move baryons through and out of galaxy disks, and are crucial to galaxy evolution by regulating star formation and enriching the circumgalactic medium \citep[CGM;][]{Rubin2014, Tumlinson2017, Veilleux2020}. Winds driven by stellar feedback are ubiquitous in high-star-formation-rate-density, high-pressure environments, which were common in cosmic noon galaxies ($z \sim 2$; \citealt{Tacconi2020}) and today exist in local starbursts. Starburst-driven winds are multiphase, as supernovae and stellar feedback generate the hot ($T\sim 10^7$~K) X-ray-emitting wind, which expands and entrains cold ($T\lesssim 10^4$~K) gas and dust. The interface between the hot wind and colder gas produces a warm/cool ($T \sim 10^5$~K) mixing layer of ionized material \citep{Heckman1990, Thompson2024}. Questions still remain about how cold gas and dust survive this hot-cold interplay as entrained clouds accelerate to the CGM. Radiative cooling \citep[e.g.,][]{Gronke2018, Gronke2020, Fielding2022} and acceleration by cosmic rays and magnetic fields \citep{Girichidis2018,Sparre2020, Rathjen2021, Armillotta2024, Sike2025} are possible solutions.

Dust is a prominent component of galactic winds. The smallest interstellar dust grains are polycyclic aromatic hydrocarbons (PAHs), which account for up to $\sim 20\%$ of the total IR luminosity in galaxies \citep{Smith2007} by processing absorbed starlight, and produce complex spectral emission features in the infrared (IR) \citep[e.g.,][]{Tielens2008, DraineLi2007, Chown2024, Peeters2024}. The relative strengths of PAH features, which trace charge and size of PAHs, depend on the intensity and shape of the radiation field \citep{Draine2021, RichieHensley2026, Baron2025} and on environmental properties such as gas density, temperature, and metallicity \citep[e.g.,][]{Micelotta2010hotgas, Aniano2020, Xu2023, Whitcomb2024, Whitcomb2025}. 

In the Spitzer era, PAHs emerged as a tracer of cooler interstellar medium (ISM) phases \citep[e.g.,][]{Regan2006, Bendo2008, Beirao2008}. Because PAHs cannot survive the initial shock and hot gas of galactic winds \citep{Micelotta2010hotgas, Micelotta2010shocks}, they trace the cool component \citep[e.g.,][]{Engelbracht2006, McCormick2013, Beirao2015}. With the improved spectral coverage, sensitivity, and resolution of JWST, PAHs and their band ratios are now powerful tools for observing cool material at parsec resolution in low-redshift galaxies \citep[e.g.,][]{Leroy2023, Sandstrom2023, Sutter2025, Chown2025, Chastenet2026}. For example, PAH band ratios revealed surprisingly minimal processing of the cool wind in the nearby starburst M82, with only minor changes in PAH charge and size, and no obvious change in abundance, with distance from its starburst \citep{Beirao2015, Cronin2026}. Correlations between PAH emission and CO, H\textsc{i}, and \ha\ gas suggest that PAHs are protected underneath the ionized surfaces of cold entrained clouds \citep{Bolatto2024, Fisher2025, Villanueva2025, Villanueva2025_erratum, Lopez2026, Donnan2026}. This protection may also explain observations in other galaxies where the cool phase traced by PAHs survives to the CGM \citep{Chastenet2026} even out to long distances and timescales \citep[$\sim 35$~kpc, $t_{\rm dyn} \sim 100$~Myr;][]{Veilleux2025}.

Like M82, NGC~253 is a nearby \citep[$\sim 3.5$~Mpc;][]{Rekola2005, Congiu2025} and highly inclined \citep[$i \sim 76^\circ$;][]{McCormick2013} galaxy hosting a central starburst \citep[star formation rate $\sim 3{-}5$~\Msunperyear;][]{Ott2005, Leroy2015_ngc253, Leroy2019} that drives a large-scale outflow.
Unlike the interaction-driven starburst in M82 \citep{Yun1994}, however, the NGC~253 starburst is fed by gas inflow along a bar \citep{Paglione2004}.
NGC~253 is also more massive \citep[$M_\star \sim 4\times10^{10}$~\Msun;][]{Leroy2021_phangs}, and its cool outflow is correspondingly more compact, with molecular streamers extending only $\sim 100{-}300$~pc in projection \citep{Bolatto2013, Walter2017}.
The wind is also projected against the disk in the south and lies behind it in the north, making the northern side difficult to detect at optical wavelengths \citep{Cronin2025}.
Nevertheless, the outflow has been measured in its cold molecular \citep{Sturm2011, Bolatto2013, Walter2017, Zschaechner2018, Krieger2019, Bao2024}, cold neutral atomic \citep{Boomsma2005, Lucero2015, Krips2016}, warm/cool ionized \citep{Matsubayashi2009, Westmoquette2011, Cronin2025}, and hot X-ray \citep{Strickland2000, Strickland2002, Bauer2007, Lopez2023} phases. Cold molecular and neutral atomic gas constitute the bulk of the mass, as is typical for galactic winds \citep{Leroy2015, Krieger2019}.

In NGC~253, dust has been detected in the nuclear starburst \citep{Dudley1999, ForsterSchreiber2003_ngc253} and halo \citep[$\sim 6{-}9$~kpc;][]{Kaneda2009}, and may correlate with H$_2$ in the molecular outflow \citep{Sugai2003}. Using ground-based VLT spectro-imaging, \citet{TacconiGarman2005} detected PAHs emitting at \three\ in the wind out to $\sim 100$~pc off-disk, with a morphology similar to the \ha\ outflow shell. 
Subsequent Spitzer observations confirmed that \seven-emitting PAHs are also entrained \citep{McCormick2013}. These studies demonstrated that small carbonaceous material can survive the hot wind of NGC~253.

In this Letter, we present the first spatially resolved study of PAH evolution along the NGC~253 outflow out to ${\sim}0.5$~kpc. Using JWST imaging, we leverage band ratios to track changes in PAH charge and size, and interpret these in the context of the entrainment and evolution of cool wind material. 
Here we adopt an infrared nucleus coordinate in the ICRS frame of $\mathrm{RA}=00^{\mathrm{h}}47^{\mathrm{m}}32.99^{\mathrm{s}}$, $\mathrm{Dec.}=-25\arcdeg17\arcmin19.74\arcsec$ \citep{Cohen2020}.

\begin{figure*}
    \centering
    \includegraphics[scale=0.65]{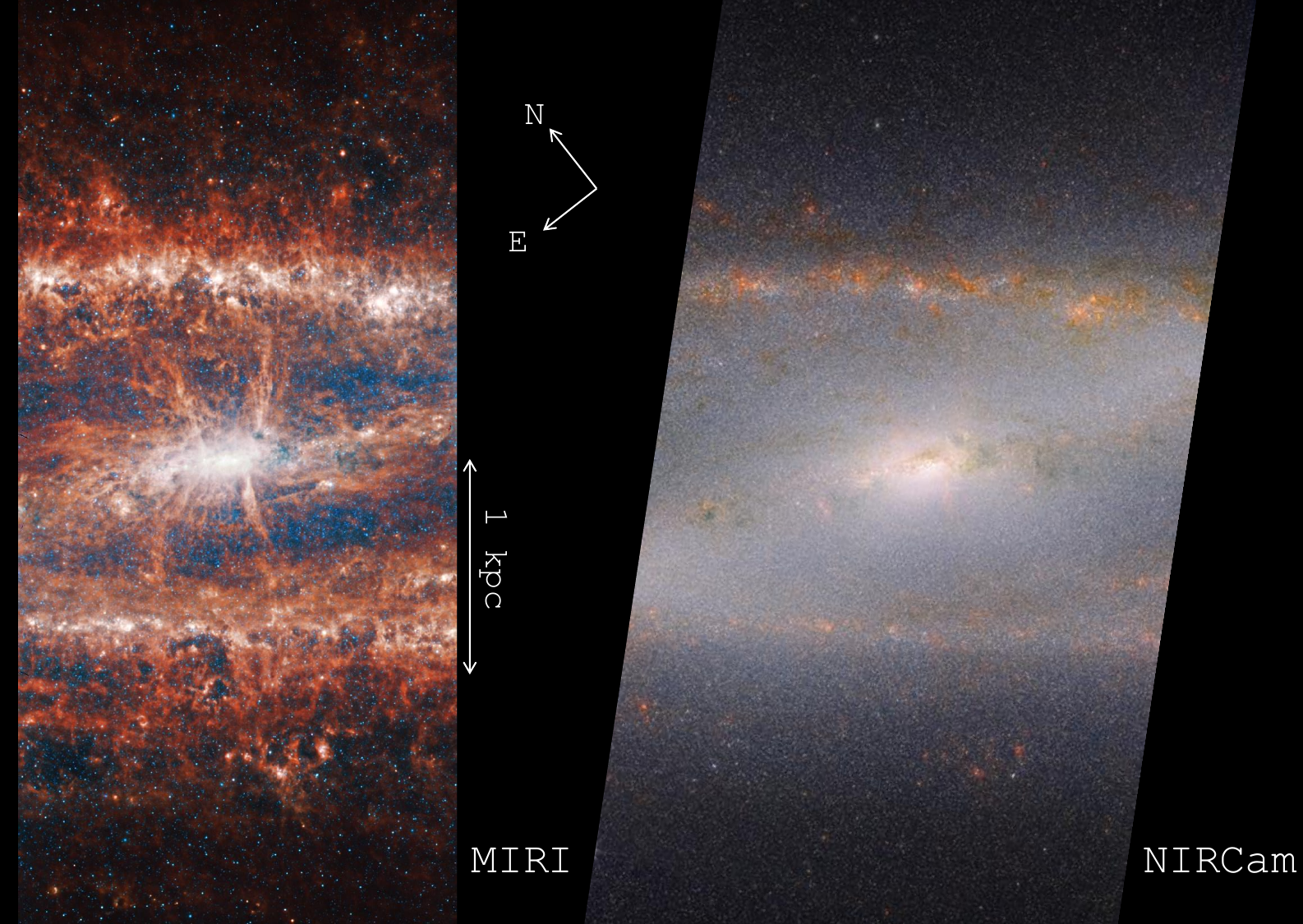}
    \caption{JWST imaging along the minor axis of NGC~253 out to $\sim \pm 2$~kpc and on $\sim 0.85{-}6.4$~pc scales. (\textit{Left}) Multi-color MIRI image using the filters F560W (cyan; continuum and some PAHs at $6.2~\mu$m), F770W (orange; PAHs at \seven\ and some continuum), and F1130W (red; PAHs at \eleven\ and some continuum). The limb-brightened edges of the outflow cone are evident, with wispy sub-structure in all four streamers. The northern wind is seen through the disk while the southern wind is pointed toward the observer and in front of the disk. (\textit{Right}) Composite NIRCam imaging using the following filters: F140M and F164N (blue; continuum and [Fe II]), F212N and F250M (green; H$_{2}$ and continuum), and F335M and F360M (red; PAHs at \three\ plus continuum). The \three\ PAH complex is brightest in the spiral arms and bar; the fainter wind is hidden by the bright stellar disk. Image processing by A.~Pagan (STScI).}
    \label{fig:pretty_pictures}
\end{figure*}

\section{Data}\label{sec:data}

Figure~\ref{fig:pretty_pictures} presents multicolor NIRCam and MIRI images of the NGC~253 wind and disk along the minor axis (GO \#1701; PI: A.~Bolatto).\footnote{Other versions of the MIRI image can be found in Fig.~1 of \citet{Thompson2024} and \citet{Donnan2026}.} At $d \sim 3.5$~Mpc, the scales probed are $0\farcs05–0\farcs375 \approx 0.85–6.4$ pc. The observing setup mirrors that of M82 in \citet[Section 2]{Cronin2026}.

\subsection{NIRCam Observations \& Data Reduction}
\begin{figure*}
    \centering
    \includegraphics[scale=0.5]{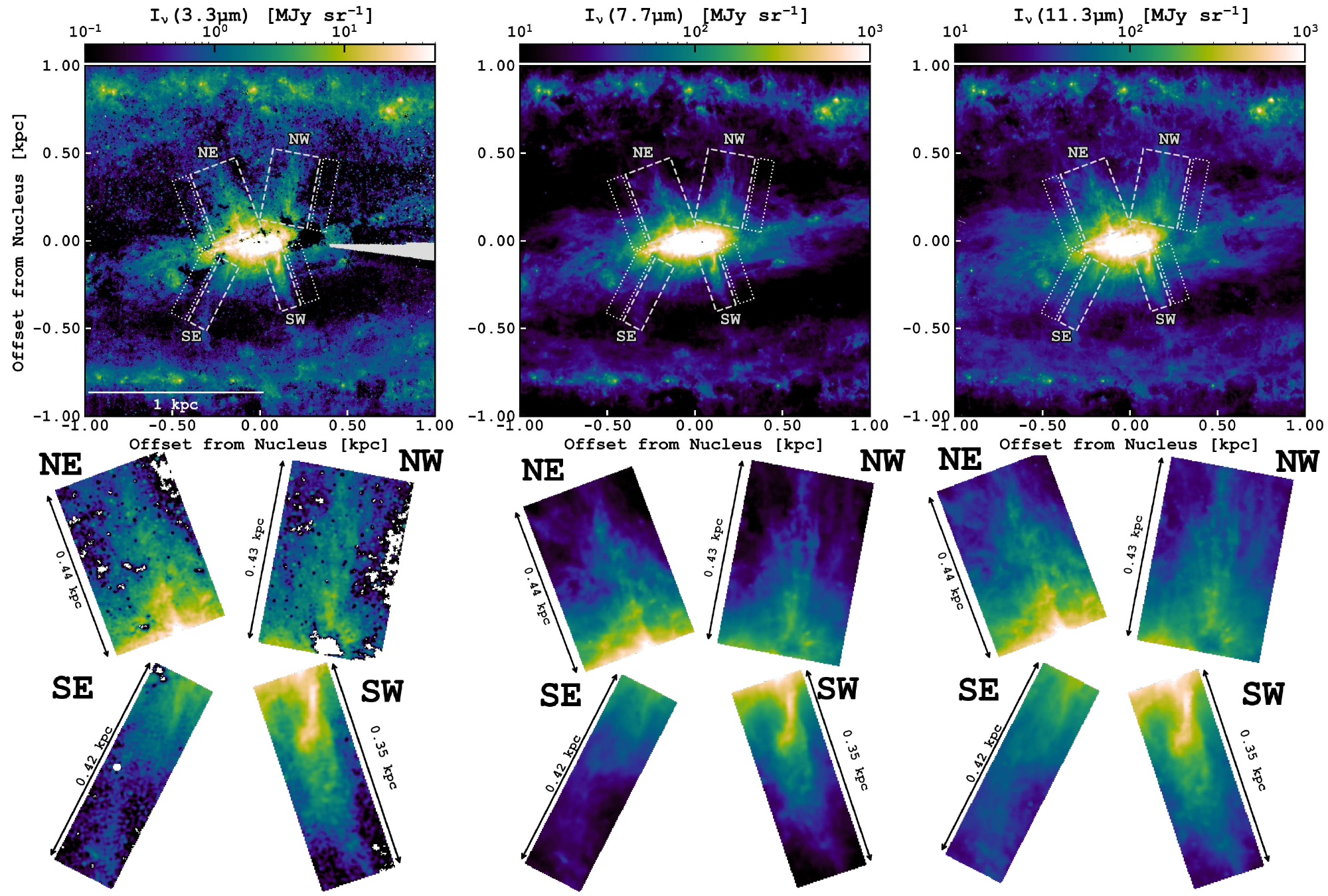}
    \caption{Emission maps of the \three, \seven, and \eleven\ PAH features in the central $2\times2$~kpc region of NGC~253. The \three\ map is continuum-subtracted from F335M using two flanking filters (F250M and F360M), while the \seven\ and \eleven\ maps are both background- and continuum-subtracted. The continuum images at all three wavelengths are in Figure~\ref{fig:continuum} of the Appendix. Labeled dashed rectangles mark the four wind streamers where we extract measurements (zoom-ins in the bottom row to show detailed sub-structure). Unlabeled dotted rectangles mark the adjacent off-streamer regions used to estimate disk-only PAH emission, since the disk contributes to each line of sight.}
    \label{fig:emission_maps}
\end{figure*}

We targeted the \three\ PAH complex using the NIRCam filters F250M, F335M, and F360M (Figure~\ref{fig:pretty_pictures}). The observations were split into a quick (401.84 s) \texttt{SUB640} exposure over the central starburst to avoid saturation, and a longer (2447.98 s) $2\times1$ \texttt{FULL} array mosaic over the remaining field of view. An incorrect PA restriction led to non-overlapping tiles in the F335M mosaic (see the \three\ map in Figure~\ref{fig:emission_maps}). We ran the JWST Calibration Pipeline \citep[version 1.17.1;][]{JWSTPipeline_1.17.1} on these observations, tweaking default parameters to better handle cosmic ray effects and saturated pixels, and incorporated the \citet{Willott2022} strategy to reduce $1/f$ noise striping. Due to a crowded field, the standard pipeline alignment was unreliable. We instead built a stellar catalog using \texttt{DOLPHOT} and aligned with the JWST/HST Alignment Tool (\texttt{JHAT}). The final mosaics contain sharp point sources and only minor artifacts from mismatched backgrounds between exposures in the image outskirts.

Using the convolution kernels from \citet{Cronin2026}, we convolved the F250M and F335M images to the resolution of F360M (PSF = $0\farcs12 \approx 2$~pc). These filters target the \three\ PAH complex, which mostly lies in the F335M bandpass but also contaminates F360M \citep{Koziol2026}. To remove the continuum, we followed the \citet{Bolatto2024} approach, which iteratively removes continuum from F335M until convergence by assuming some fraction of PAH contamination in F360M and a completely PAH-free F250M. We set this fraction to $0.55$ following \citet{Cronin2026}. 
This process also assumes a value that represents where the F335M continuum lies between the flanking F250M and F360M filters. We find that $0.75$ mitigates over-subtraction in the outflow streamers, and is close to the central wavelength ratio $(3.35~\micron - 2.5~\micron) / (3.6~\micron - 2.5~\micron) \approx 0.77$. Although this value over-subtracts some regions of the disk, Figure~\ref{fig:continuum} in the Appendix shows that the resulting continuum image contains no obvious artifacts. The median continuum contribution to F335M is $\sim 90{-}95\%$.

Finally, we convolved the continuum-subtracted \three\ image to the resolution of F1130W (PSF = $0\farcs375 \approx 6.4$~pc), and present it in Figure~\ref{fig:emission_maps}.

\subsection{MIRI Observations \& Data Reduction}
We targeted the MIRI filters F560W, F770W, and F1130W. The latter two capture the \seven\ and \eleven\ PAH complexes, while F560W contains continuum and a partial \six\ PAH feature. The MIRI observations were taken with a short (9.52 s) \texttt{SUB128} exposure on the bright starburst and a longer (2664.03 s) $1\times4$ mosaic \texttt{FULL} array exposure along the minor axis. 
A single tile \texttt{FULL} background exposure was obtained off-galaxy at (R.A., Dec.) J2000 = ($12.0055^\circ$, $-25.3812^\circ$).

We used version 1.12.5 of the JWST Calibration Pipeline \citep{JWSTPipeline_1.12.5}, keeping default parameters except for F770W, where we retain the first group in each integration to recover saturated pixels in the central starburst. We then subtracted a median background (mostly from instrumental thermal emission and zodiacal light) from the F560W ($1.4$~MJy~sr$^{-1}$), F770W ($5.4$~MJy~sr$^{-1}$), and F1130W ($22.4$~MJy~sr$^{-1}$) images, estimated from the background exposure. The background-subtracted F560W and F770W images were matched to the resolution of F1130W (PSF = $0\farcs375 \approx 6.4$~pc) using the \citet{Cronin2026} kernels.

To remove the underlying dust continuum from the \seven\ and \eleven\ PAH complexes, we followed the procedure of \LHands, who performed synthetic JWST photometry on spectroscopy of galaxies in the Spitzer IRS Mapping Legacy Archive \citep{Donnelly2025_SIMLA}. \LHands\ fit functional forms to map the photometric color to the percent contribution of PAH emission to the F770W and F1130W filters \citep[from \texttt{PAHFIT} spectral decomposition;][]{Smith2007} versus MIRI color (F1130W/F560W and F770W/F560W) diagram. We applied these functions to our F770W and F1130W images, resulting in the continuum-subtracted maps in Figure~\ref{fig:emission_maps}. The typical continuum contribution is $\sim 20\%$ to both filters (see Figure~\ref{fig:continuum}), consistent with the continuum percentages calculated from Spitzer spectroscopy in M82 \citep{Lopez2026, Cronin2026}.

\begin{figure*}
    \centering
    \includegraphics[scale=0.48]{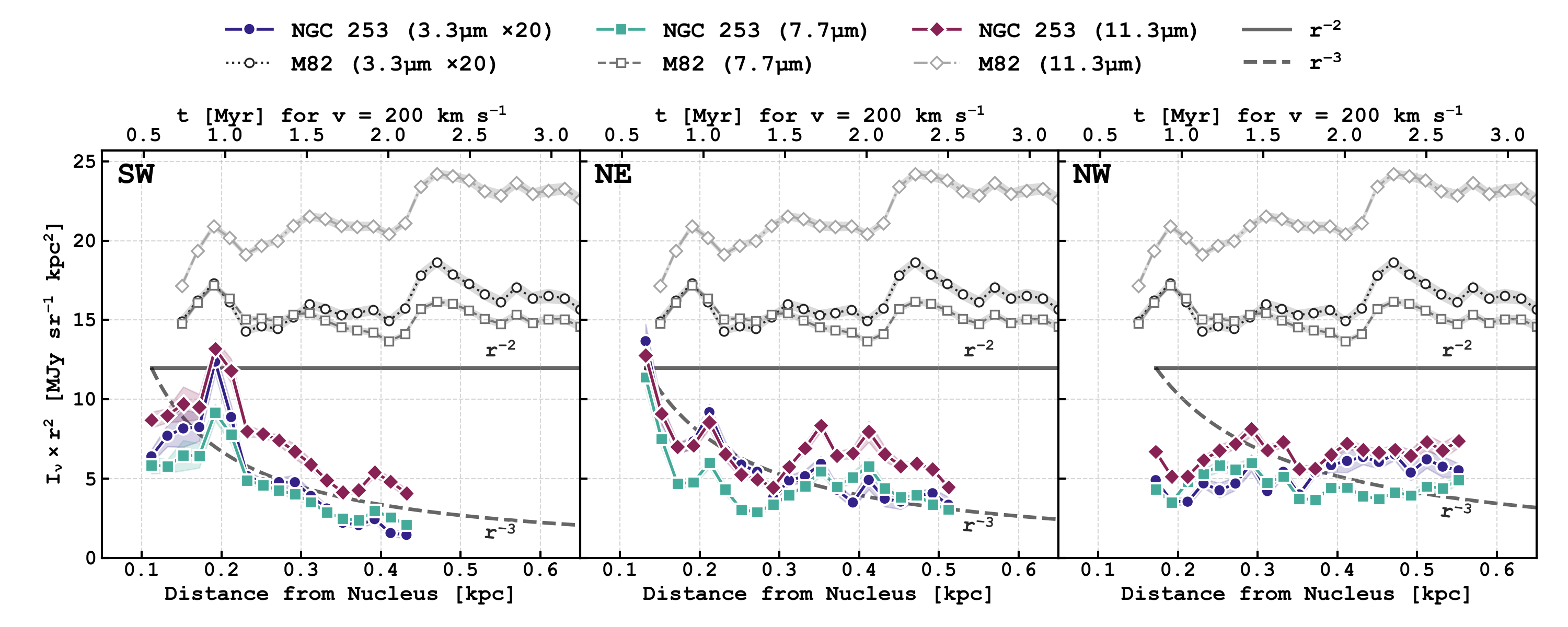}
    \caption{Binned-mean emission profiles extracted along each NGC~253 streamer (colored points) with M82 (gray points) and arbitrary $r^{-2}$ and $r^{-3}$ profiles for comparison. The shaded regions are the standard error of the mean.
    Each profile is multiplied by $r^{2}$ to divide out the geometric dilution expected for illumination by the starburst, so that a flat ($r^{-2}$) profile indicates changes in PAH emission due to distance from the starburst. The top $x{-}$axis converts the deprojected distance to a timescale assuming a cool-outflow speed of $200$~\kms. All PAH bands in the SW and NE streamers decline faster than $r^{-2}$, suggesting possible changes to PAH emission that are not due to distance from the illuminating starburst. The NW streamer roughly follows $\sim r^{-2}$ much like M82. We exclude the SE streamer, where we do not produce a reliable disk subtraction.}
    \label{fig:emission_profiles}
\end{figure*}

\section{Results \& Discussion} \label{sec:results}
\subsection{PAH Emission Along the Outflow Streamers}

Figures~\ref{fig:pretty_pictures} and \ref{fig:emission_maps} provide the highest-resolution IR view of the NGC~253 wind to date. JWST detects all four limb-brightened edges of the outflow cone, on both sides of the disk, at \three, \seven, and \eleven\ out to a deprojected distance $r \sim 0.5$~kpc. These streamers correspond to the locations of the molecular streamers identified in CO \citep{Bolatto2013} which JWST now resolves into filamentary sub-structure ($\sim 0.85{-}6.4$~pc resolution). Outflow emission is also detected between the edges of the wind cones, including a possible arc in the north that resembles a scaled-down version of the prominent arcs seen in M82 \citep{Lopez2025, Cronin2026}. We focus our analysis on the streamers because they stand out clearly against the intervening disk, and they are well-studied in the literature \citep[e.g.,][]{Heesen2011, Bolatto2013, Walter2017, Zschaechner2018, Levy2022, Bao2024}. 

We produce radial profiles of the PAH emission along the streamers by averaging the brightest 25\% of pixels across each dashed rectangle in Figure~\ref{fig:emission_maps}. We determine the rectangle geometry by eye, favoring larger widths to capture the full extent of each streamer, and selecting only the brightest pixels to ensure that we are not including adjacent disk-only emission. Future work could employ more sophisticated methods to isolate the streamers. 

Because the outflow is in projection with the disk, each line of sight along the streamers contains a mixture of outflow and disk emission. We estimate disk emission as the mean of the faintest 25\% of pixels in the dotted rectangles adjacent to each streamer, and then subtract this mean from the total streamer brightness to produce outflow-only profiles. The disk contributes a median of $\sim$4--14\% of the streamer brightness at \three\ and $\sim$30--40\% at \seven\ and \eleven. The overall radial trends down each streamer are consistent when we do not subtract the disk, indicating that these trends are driven by the streamer emission. The southeast (SE) streamer, however, has low contrast with the surrounding disk material and thus appears over-subtracted in our profiles regardless of where we estimate the disk-only emission, and so we exclude it from our analysis. Additionally, the outflow-only profiles turn over at $r \sim 0.5$~kpc in the southwest (SW) and northeast (NE) and $r \sim 0.6$~kpc in the northwest (NW), where they begin to increase due to confusion with the spiral arms. We truncate the profiles before this turnover so that they trace only outflow emission.

\begin{figure*}
    \centering
    \includegraphics[scale=0.48]{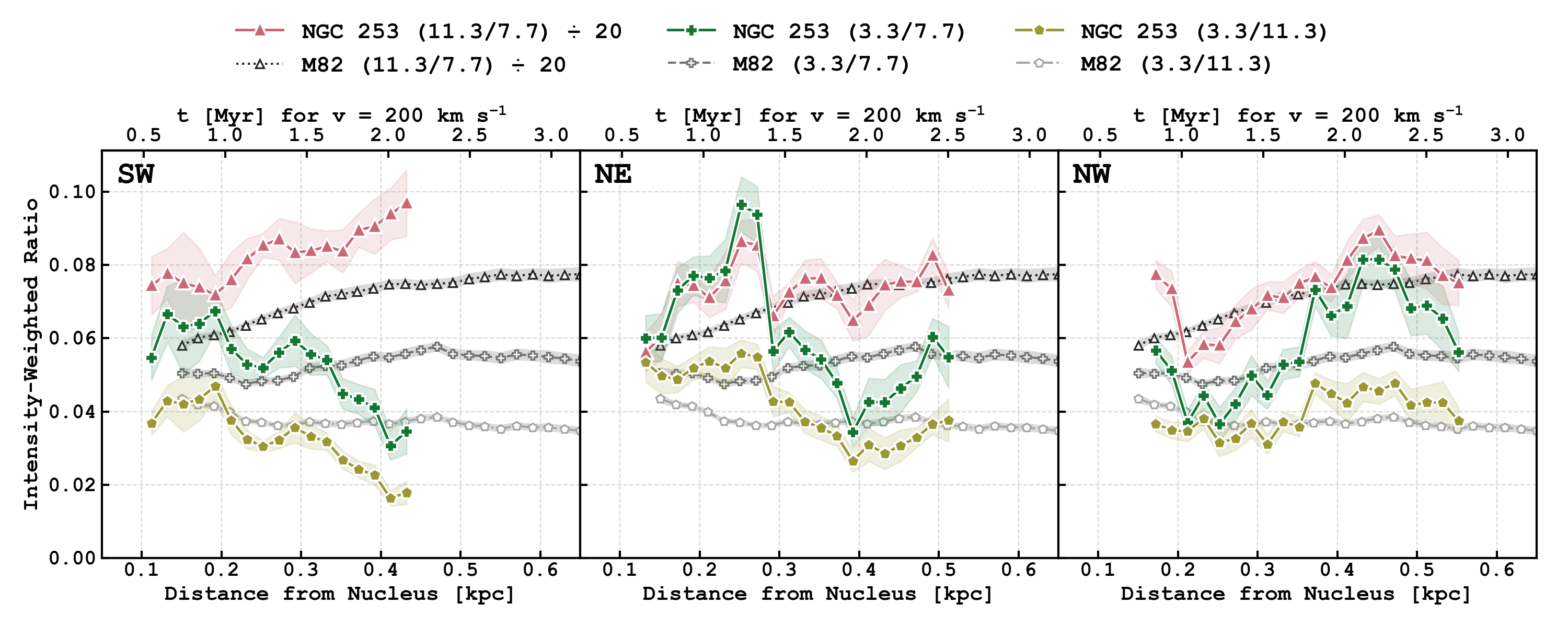}
    \caption{PAH ratios along each NGC~253 streamer, computed from the emission profiles in Figure~\ref{fig:emission_profiles}, with the M82 ratios for comparison \citep[gray;][]{Cronin2026}. Ratios are intensity-weighted so that fainter regions do not skew them. We divide \ElevenSeven\ by 20 for display, and the $x{-}$axes are deprojected distance from the starburst nucleus and the timescale over that distance assuming $v \sim 200$~\kms. The SW streamer shows clear trends of the \ElevenSeven\ ratio increasing with distance, as in M82, and the \ThreeSeven\ and \ThreeEleven\ ratios decreasing, unlike the roughly flat ratios of M82. Beyond ${\sim}0.2{-}0.3$~kpc, \ElevenSeven\ in the NE and NW agrees with M82. The \three\ ratios vary from streamer to streamer and do not follow M82’s flat profiles.}
    \label{fig:ratio_profiles}
\end{figure*}

In Figure~\ref{fig:emission_profiles}, we plot the binned mean of the outflow-only radial profiles, with uncertainties given as the standard error of the mean\footnote{i.e., the standard deviation divided by the square root of the number of beams, added in quadrature for the streamer and adjacent rectangle.}, alongside binned-mean M82 profiles from \citet{Cronin2026}. The radial distance is deprojected assuming $i = 76^\circ$ and that the wind emerges from the plane of the disk at a semi-opening angle of $\alpha = 30^{\circ}$ measured from the cone axis \citep{Westmoquette2011}. This geometry also makes the timescale over which these PAHs evolve uncertain. The molecular outflow has a deprojected velocity of $v_{\rm deproj} \sim 200{-}400$~\kms\ \citep{Bolatto2013, Walter2017, Bao2024}, corresponding to $t_{\rm dyn} \sim 1{-}2.5$~Myr along the $r \sim 0.5$~kpc deprojected length of the SW streamer, although the outflow may be accelerating \citep{Walter2017, Cronin2025}. In Figures~\ref{fig:emission_profiles} and \ref{fig:ratio_profiles} and throughout this Letter, we adopt $v_{\rm deproj} \sim 200$~\kms\ for consistency with \citet{Cronin2026} on M82, which is also in line with dust speeds in simulations of starburst-driven outflows \citep{Richie2026} and with possible outflow velocities of the \eleven\ PAH feature toward the center of NGC~253 \citep{Donnan2026}. Similar velocities are measured for the molecular and ionized gas in the M82 wind \citep{Shopbell1998, Krieger2019, Leroy2015}, corresponding to $t_{\rm dyn} \sim 5{-}10$~Myr over $\sim 2$~kpc \citep{Villanueva2025,Villanueva2025_erratum,Lopez2026, Cronin2026}. If we were to assume speeds of the PAH-bearing phase to be closer to the \ha\ velocity ($\sim 500$~\kms; \citealt{Cronin2025}), these timescales would drop by a factor of $\sim 2.5$ at most.

In Figure~\ref{fig:emission_profiles}, we multiply the intensity profiles by $r^2$ to reveal possible variations in emission that do not depend on distance from the illuminating starburst. This assumes that flux declines as $r^{-2}$, where $r$ is the deprojected radial distance from an illuminating point source, and a uniformly illuminated outflow cone. This approximation breaks down for wind material close to the starburst, which is spatially extended, but improves at larger $r$ where the starburst subtends a smaller solid angle. Radial profiles that deviate from $r^{-2}$ indicate changes in emission that cannot be solely attributed to distance from the starburst. For reference, we include arbitrary $r^{-2}$ and $r^{-3}$ profiles in Figure~\ref{fig:emission_profiles}. We find that the SW streamer declines more steeply than $r^{-2}$ in all three bands, as $r^{-3.2}$, $r^{-3.1}$, and $r^{-2.8}$ at \three, \seven, and \eleven, respectively. The NE streamer declines as $r^{-2.8}$, $r^{-2.4}$, and $r^{-2.3}$, and the NW streamer as $r^{-1.6}$, $r^{-2.1}$, and $r^{-1.9}$, in the same bands, with typical uncertainties on the power-law indices of $\pm0.1$--$0.2$.

Assuming stochastic heating, PAH intensity is expected to scale as $I_{\rm PAH} \propto U \times N({\rm H}) \times {\rm DGR} \times q_{\rm PAH}$, where $U$ is the radiation field intensity, $N({\rm H})$ is the column density of PAH-bearing gas, DGR is the dust-to-gas ratio, and \qpah\ is the fraction of dust mass found in PAHs \citep{DraineLi2007}. A steep ($\sim r^{-3}$) radial decline along the SW streamer in all bands and the NE streamer at \three\ therefore indicates that $N({\rm H})$, DGR, or \qpah\ decline with distance, or that $U$ declines faster than geometric dilution. Assuming a steady-state wind, this decline may hint at PAH destruction in the outflow, i.e., a declining \qpah. The bump around $0.2$~kpc in the SW streamer could instead trace a recent outflow event, with newly ejected material still close to the disk, i.e., a varying $N({\rm H})$. These results differ from M82, where radial variations in PAH intensity appear to be mostly governed by distance from the illuminating starburst \citep{Cronin2026}. The PAH-to-gas profiles are relatively flat in M82 \citep{Lopez2026}, suggesting either that PAHs are protected in dense gas or that the DGR does not change. Beyond these factors, PAHs in the M82 and NGC~253 outflows may be exposed to different radiation field shapes and strengths.

\begin{figure*}
    \centering
    \includegraphics[scale=0.48]{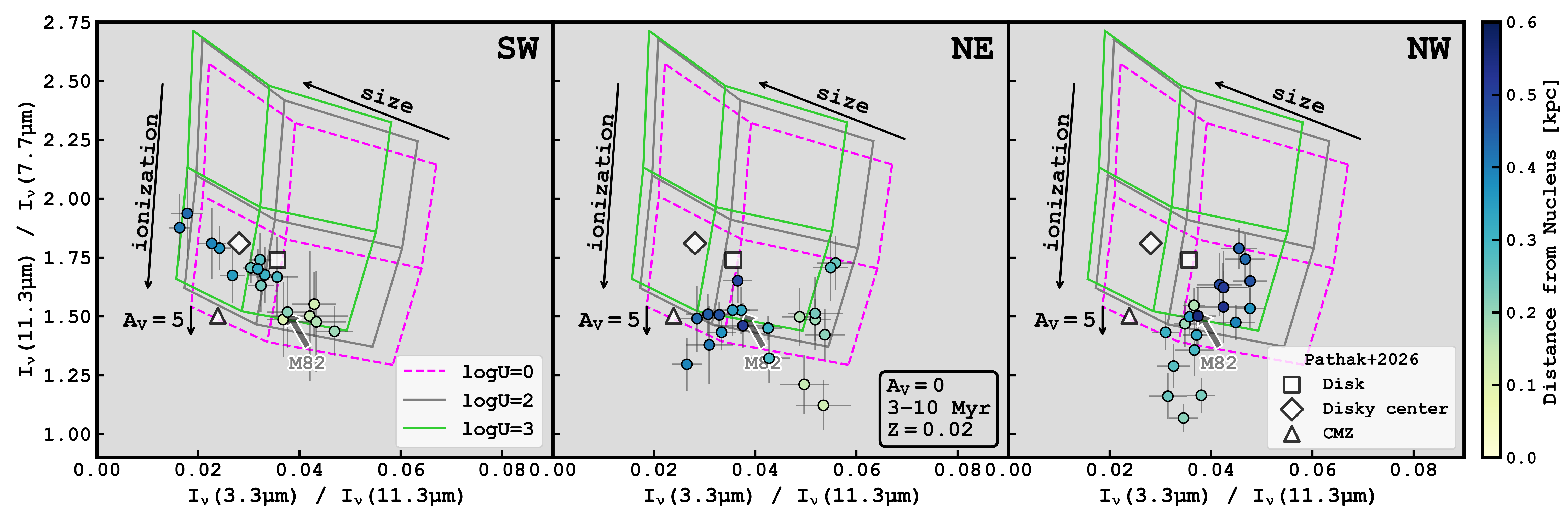}
    \caption{Ratios from Figure~\ref{fig:ratio_profiles} compared with continuum-subtracted \citet{Draine2021} dust models, assuming a $3{-}10$~Myr starburst at solar metallicity \citep{BC03}. Each grid corresponds to a radiation field intensity $U$ at $A_{\rm V} = 0$. The small black arrow shows the magnitude and direction of how the grids shift for a higher extinction of $A_{\rm V} = 5$, and the long black arrows show the effects of PAH size and ionization. Points are colored by distance from the nucleus. The gray arrow shows the M82 trend from $0.3$~kpc (excluding inner points affected by silicate absorption) to $0.6$~kpc (the most extended NGC~253 streamer). White symbols show median PHANGS-JWST ratios across galaxies \citep{Pathak2026} for star-forming disks, disk-like galaxy centers (``disky center''), and young star-forming centers or central molecular zones (CMZs). Silicate absorption affecting the \eleven\ feature near the starburst may explain some ratios outside the $A_{\rm V} = 0$ grids. In the SW streamer, PAHs become larger and more neutral with distance, matching M82 but spanning a wider range over the same distance and timescale. NGC~253 ratios fall near the PHANGS-JWST medians, with the outermost SW and NE ratios matching disks and disk-like centers, and the innermost ratios matching CMZs. Figure~\ref{fig:draine_models_no_disksub} shows the NGC~253 ratios without disk subtraction.}
    \label{fig:draine_models}
\end{figure*}

\subsection{Evidence of PAH Processing with Band Ratios}
To understand whether the properties of the PAHs themselves are also changing along the outflow, we take ratios between band intensities, which are sensitive to PAH size and charge \citep{Tielens2008,Rigopoulou2021,Draine2021} and also radiation field hardness \citep{Draine2021, RichieHensley2026}. Higher values of \ElevenSeven\ indicate neutral and possibly larger PAHs. Because the \three\ feature traces the smallest PAHs, larger \ThreeSeven\ and \ThreeEleven\ suggest a smaller size distribution, though \ThreeSeven\ is also sensitive to charge, and \three\ emission responds most strongly to radiation field hardness as the smallest PAHs \citep[$N_{\rm C} \lesssim 100$ carbon atoms;][]{Draine2021} have the lowest heat capacities. Tracking variations in these ratios throughout the wind helps us infer the possible evolution of PAH-bearing clouds.

Using the emission profiles from Figure~\ref{fig:emission_profiles}, we calculate PAH ratios along the SW, NE, and NW streamers and present them in Figure~\ref{fig:ratio_profiles}. The ratios are weighted by intensity to down-weight regions of low emission, and we include the M82 ratios over the same distance for comparison. The \ElevenSeven\ ratios are divided by 20 for display. 
Radial profiles of the ratios without disk subtraction are in the Appendix (Figure~\ref{fig:ratio_profiles_no_disksub}).

The SW streamer shows the clearest trends. The \ElevenSeven\ ratio increases ($0.32\pm0.09$~dex~kpc$^{-1}$), while the \ThreeSeven\ and \ThreeEleven\ ratios decline ($-0.72 \pm 0.15$ and $-1.00 \pm 0.18$~dex~kpc$^{-1}$) with distance from the starburst. For \ElevenSeven, this trend matches M82 over comparable distances and timescales (assuming a similar deprojected cool-phase velocity of $\sim 200$~\kms); the NGC~253 \ThreeSeven\ and \ThreeEleven\ trends, however, differ from the mostly constant ratios found in M82, possibly suggesting a decrease in small PAHs in NGC~253 that is not seen in M82. This decline in the \three\ ratios is also seen in the NE streamer, where the net change is $-0.61 \pm 0.19$ and $-0.73 \pm 0.12$~dex~kpc$^{-1}$ for \ThreeSeven\ and \ThreeEleven, respectively, while \ElevenSeven\ is flat ($0.09 \pm 0.07$~dex~kpc$^{-1}$). In the NW, \ElevenSeven\ increases ($0.34 \pm 0.10$~dex~kpc$^{-1}$) much like the SW streamer and M82, but differs from all streamers and M82 with increasing \ThreeSeven\ ($0.61 \pm 0.16$~dex~kpc$^{-1}$) and \ThreeEleven\ ($0.33 \pm 0.09$~dex~kpc$^{-1}$). The shapes of these radial profiles hold for the uncorrected (disk+outflow) ratios (Figure~\ref{fig:ratio_profiles_no_disksub}), meaning the overall trends are likely driven by outflow emission and not disk. It remains unclear why the ratio profiles between the SW and NE streamers match well while the NW deviates for the \three\ ratios.

Comparison with dust models provides more robust insight into the significance of these varying PAH band ratios. Using \texttt{synphot} and the JWST throughput curves \citep{Glasse2015}, we perform synthetic photometry on model spectra from \citet{Draine2021} to produce synthetic photometric bands for F250M, F335M, F360M, F770W, and F1130W. Following \citet{Cronin2026}, the model spectra assume a \citet{BC03} $3{-}10$~Myr old starburst radiation field at solar metallicity. With F250M, F335M, and F360M, we repeat the \three\ continuum subtraction process from Section~\ref{sec:data}. For the synthetic MIRI bands, we subtract the median continuum contribution to each band in the outflow streamers ($\sim 20\%$; Figure~\ref{fig:continuum}), producing grids in the \ElevenSeven${-}$\ThreeEleven\ plane (Figure~\ref{fig:draine_models}). Each grid represents a different radiation field heating the dust ($U = 1, 10^2, 10^3$, chosen assuming a $r^{-2}$ dilution of the starburst luminosity \citealt{Engelbracht1998}). The long black arrows show the effect of PAH size and ionization, informed by \citet{Draine2021}, and the short black arrow shows the direction and magnitude of how the grids shift when assuming $A_{\rm V} = 5$, computed using a \cite{Gordon2023} extinction curve with $R_V = 3.1$ \citep{Gordon2024}, which includes the 9.7 µm silicate feature at average Milky Way values. For reference, \citet{Cronin2025} used the Balmer decrement to find $A_{\rm V} \gtrsim 5$ in the nucleus of NGC~253 and $A_{\rm V} \sim 1$ in the southern outflow, and \citet{McClain2026} measured $A_{\rm V} \sim 3$ in the brightest \ion{H}{2} regions throughout the disk.

The points in Figure~\ref{fig:draine_models} are colored by distance from the nucleus, and the gray arrow shows the M82 trend from \citet{Cronin2026} between $0.3{-}0.6$~kpc, excluding inner radii affected by silicate absorption, and matching the furthest streamer extent in NGC~253. For comparison with typical star-forming galaxies, we include median PHANGS-JWST ratios for star forming disks (square), disk-like galaxy centers (``disky center;'' diamond), and young star-forming centers or central molecular zones (CMZs; triangle) from \citet{Pathak2026}. A version of this plot without the disk-subtracted ratios are in the Appendix (Figure~\ref{fig:draine_models_no_disksub}).

Overall, we find good agreement between the model grids and the NGC~253 ratios, especially in the SW streamer. The points falling below the grids in the NE and NW streamers may be better constrained with models including higher extinction (see the $A_{\rm V} = 5$ arrow), possibly due to silicate absorption at $9.7~\mu$m affecting the \eleven\ feature, which has been found to severely affect mid-IR features in the starburst of M82 with a depth of $\tau_{\rm 9.7} \sim 1.5$ \citep{Cronin2026, Duval2026}, consistent with the depth found in the starburst of NGC~253 with MIRI (K.~Donaghue~et al.~in prep). Even so, there are significant variations in the ratios at further distances that may not be as affected by silicate absorption. The SW streamer shows a clear trend toward more neutral and larger grains, with size showing the largest magnitude of change. In the NE streamer, PAHs shift toward larger grains in the extended wind, but a trend with ionization is less clear in the disk-subtracted ratios. However, the uncorrected ratios in Figure~\ref{fig:draine_models_no_disksub} reveal a slight shift toward more neutral PAHs with distance, meaning that the disk subtraction has added scatter in Figure~\ref{fig:draine_models}. PAHs in the NW streamer become more neutral with distance, but neither the disk-subtracted nor uncorrected ratios show a clear size trend. Possible over-subtraction of the \three\ feature $\lesssim 0.2$~kpc from the nucleus (see the black patch in Figure~\ref{fig:emission_maps}) may be responsible. We also find agreement between the NGC~253 wind ratios and the typical galaxy measurements from PHANGS-JWST \citep{Pathak2026}. The outermost SW and NE ratios match the values for disk and disk-like centers, while the innermost ratios match those of star-forming centers and CMZs.

PAHs becoming more neutral as they are accelerated along the SW, NW, and possibly NE streamers is a key overlap with the results of M82. \citet{Cronin2026} argue that this change in ionization is because PAHs are exposed to a weaker radiation field and ionization parameter ($\gamma$) as they are transported away from the starburst. However, the PAH size distribution differs between both winds. In M82, most of the wind ratios cluster around the  standard-to-large size of \citet{Draine2021} dust models \citep[see Figure.~7 of][]{Cronin2026}. In Figures~\ref{fig:draine_models} and \ref{fig:draine_models_no_disksub}, the gray arrow shows how the M82 points change over a similar distance and timescale as NGC~253. While PAHs become larger and more neutral with distance, much like the SW and NE streamers, the magnitude of change in the size distribution in M82 is much smaller than in NGC~253. This may suggest that NGC~253 contains smaller PAHs than M82 at shorter distances, and perhaps overall faster evolution in PAH size on a $\sim 1{-}2.5$~Myr timescale.

This observed PAH evolution in the NGC~253 wind may be due to proximity to the starburst. For instance, \qpah\ in the M82 wind is low ($\sim 1\%$), likely because supernova feedback in the starburst alters \qpah\ in the ambient ISM before it is entrained in the flow. When including the wider Spitzer imaging, \qpah\ stabilizes across $\sim 2{-}5$~kpc, or $\sim 5{-}20$~Myr \citep{Cronin2026}. This stability strongly indicates that PAHs surviving breakout are protected from the hot wind, probably cocooned under the surfaces of cool clouds, and that the destruction setting this \qpah\ has largely ceased by the time the wind reaches these radii. Additionally, the feedback in the disk may preferentially destroy smaller PAHs, as seen in AGN and stellar feedback environments \citep[e.g.,][]{Croiset2016, Lai2023, Lai2025, GarciaBernete2022a, GarciaBernete2022b, Maragkoudakis2026}, explaining why most of the M82 ratios on longer timescales ($\gtrsim 2.5$~Myr) suggest the presence of standard-to-large PAHs. Because the M82 measurements are less reliable at $r \lesssim 0.3$~kpc, the earliest phase of PAH processing in the wind (when the smallest PAHs are destroyed) may be inaccessible.

NGC~253, on the other hand, contains PAH-rich streamers that are less obscured at $r \lesssim 0.3$~kpc. Over this distance, PAHs shift from small to large (Figure~\ref{fig:draine_models}), and the $A_{\rm V} = 5$ arrow shows that extinction mainly affects charge measurements and not size. Thus, our proximity to the NGC~253 starburst may mean we are sampling an earlier phase of PAH evolution in winds when small PAHs are still being destroyed, whereas the M82 measurements probe the population that has largely survived this feedback. The clear radial trend in the NGC~253 \three\ ratios, against the flatter M82 profiles, is consistent with more size evolution over a shorter timescale.

PAHs may be destroyed at the base of the wind from a number of mechanisms, such as shocks, hot gas, or even cosmic rays, each of which can preferentially destroy smaller PAHs. PAHs with $N_{\rm C} = 50$~carbon atoms do not survive shocks faster than $\sim 100$~\kms, while those with $N_{\rm C} = 200$ can survive up to $\sim 125$~\kms\ \citep{Micelotta2010shocks}. SiO and CH$_3$OH emission reveal enhanced shock strength at the base of the outflow streamers \citep{Bao2024}, particularly the SW streamer, where we find the strongest PAH size trend. Cosmic rays also preferentially destroy smaller PAHs, though on $\gtrsim 20$~Myr timescales for $N_{\rm C} = 50$ at solar-neighborhood fluxes \citep{Micelotta2011}. However, ionization rates in the NGC~253 CMZ are $\sim 10^2{-}10^3\times$ higher, and scaling linearly would reduce this lifetime to $\lesssim 0.2$~Myr in the central $\sim 100$~pc and $0.2{-}1$~Myr beyond \citep{Behrens2024}. Cosmic rays are advected out by the outflow \citep{Heesen2011}, and $1.5$~GHz synchrotron emission extends along the PAH streamers \citep[][ Figure~\ref{fig:radio} in the Appendix]{Heesen2011}, tracing cosmic-ray electrons and shock-compressed magnetic fields. Finally, if shocks are slower than $< 100$~\kms\ and cosmic ray destruction is too long, smaller PAHs could still be preferentially destroyed in the post-shock gas, but only at $T \lesssim 10^6$~K \citep{Micelotta2010hotgas}. In the $\sim10^7$~K X-ray-emitting outflow of NGC~253 \citep{Lopez2023}, PAHs of all sizes should be destroyed within $\lesssim 10^3$~yr. We note that these interpretations assume a steady-state flow, in which distance traces time since launch; an episodic starburst would generate a more dynamic outflow rate, and thus imprint different outflow events at each radius.

\subsection{PAH-bearing Clouds Survive the Flow} \label{sec:timescales} 
The PAH-emitting streamers coincide with the hot, X-ray emitting wind \citep{Strickland2000, Strickland2002, Lopez2023}, implying that cool, dusty gas must survive within a much hotter medium. Theory suggests that for a cool cloud to survive a hot wind, radiative cooling in the cloud--wind interface must act faster than hydrodynamical instabilities can disrupt the cloud, $t_{\rm cool,mix} \ll t_{\rm cc}$, where $t_{\rm cool,mix}$ is the cooling time of the mixed interface gas \citep{Klein1994, Gronke2018, Abruzzo2022, Farber2022, Fielding2022}. The ``cloud-crushing'' timescale is $t_{\rm cc} = \chi^{1/2} r_{\rm cl} / v_{\rm rel}$, where $\chi$ is the ratio of the cloud to wind density, $r_{\rm cl}$ is the initial radius of a spherical cloud, and $v_{\rm rel}$ is the relative velocity between wind and cloud. We apply this criterion to the SW streamer, which shows the clearest trends in PAH emission and the band ratios. It is also the brightest outflow component in molecular gas, and hence the best-studied part of the NGC 253 outflow \citep[e.g.,][]{Heesen2011, Bolatto2013, Walter2017, Zschaechner2018, Levy2022, Bao2024, Cronin2025}.

\citet{Levy2022} measured beam-deconvolved FWHM $=8$~pc in $350$~GHz dust continuum and $13$~pc in CO($3-2$) for the SW streamer. We take $r_{\rm cl} \sim {\rm FWHM}/2 \approx 4-6.5$~pc, which is a characteristic scale rather than an exact cloud radius since the SW streamer could be composed of multiple clouds. Assuming $v_{\rm hot} \sim 1000$~\kms\ \citep{Lopez2023} and a PAH-emitting phase at $v_{\rm cool} \sim 200{-}400$~\kms, then $v_{\rm rel} \sim 600{-}800$~\kms. Simulations suggest $\chi \sim 100{-}1000$ \citep{Gronke2018, Sparre2020, Abruzzo2022}, yielding $t_{\rm cc} \sim 0.05{-}0.35$~Myr.

Adopting $T_{\rm cool} \sim 10^4$~K, the X-ray temperatures ($T_{\rm hot} \sim 10^7$~K) and pressures ($P/k_{\rm b} \sim 10^6$~K~cm$^{-3}$) from \citet{Lopez2023}, and a cooling curve of $\Lambda(T_{\rm mix}) \approx 10^{-21.4}$~erg~s$^{-1}$cm$^{3}$ \citep{Abruzzo2022}, we find $t_{\rm cool, mix} \sim 10^{-3}$~Myr \citep[see][for details]{Gronke2018, Abruzzo2022, Farber2022}. This satisfies the $t_{\rm cool, mix} \ll t_{\rm cc}$ criterion by about two orders of magnitude, meaning material in the SW streamer is cooling rather than being shredded in the flow, consistent with findings at the base of the M82 wind \citep{Fisher2025}. Moreover, Kelvin-Helmholtz instabilities grow to a cloud radius in about one $t_{\rm cc}$, and clouds are adiabatically mixed into the flow within a few $t_{\rm cc}$ \citep{Klein1994}. Given $t_{\rm dyn} \sim 1{-}2.5$~Myr, the SW streamer has undergone $\sim3{-}50~t_{\rm cc}$.\footnote{If we were to instead adopt $v_{\rm cool} \sim 500$~\kms\ used by \citet{Fisher2025} at the base of the M82 wind, the dynamical time becomes $t \sim 1$~Myr, yielding $\sim2{-}13~t_{\rm cc}$ \citep[see also the discussion in ][]{Cronin2026}.} In essence, the SW streamer has persisted for at least two to three cloud-crushing times given every combination of $\chi$, $r_{\rm cl}$, and $v_{\rm cool}$ considered here.

However, the JWST images in Figure~\ref{fig:pretty_pictures} reveal that the cool wind, as traced by PAHs, extends only to $r \sim 0.5$~kpc due to a combination of galaxy mass and line-of-sight confusion with the spiral arms. NGC~253 is a more massive galaxy than M82 \citep[$4\times10^{10}$~\Msun~vs.~$5\times10^9$~\Msun][]{Lehmer2008, Leroy2021_phangs} and thus has a deeper gravitational potential, making it more difficult for material to escape. Indeed, while the \ha\ and X-ray phases extend out to $\sim 1{-}2$~kpc \citep{Westmoquette2011, Lopez2023, Cronin2025} and $\sim 10$~kpc \citep{Strickland2000, Strickland2002}, respectively, the slower-moving molecular phase only reaches a few hundred parsecs \citep{Bolatto2013, Walter2017}, much like the PAH-rich phase measured here. 
What becomes of PAHs surviving breakout is unclear. Because they are likely protected in cold clouds, their radial extent may simply match the molecular gas, and thus be mainly influenced by the gas column density; alternatively, beyond $r \gtrsim 0.5$~kpc the lines of sight may be confused, where wind emission becomes faint and difficult to separate from the PAH-rich spiral arms. A constraint on the PAH abundance \qpah\ along the outflow streamers would help us distinguish these effects from actual PAH destruction in the flow, assuming an unchanging DGR, but for now is difficult because current \qpah\ estimates require far-IR information from Herschel to constrain \LTIR\ (\qpah\ $\propto$ \LPAH/\LTIR). Herschel's non-axisymmetric $\sim 6''$ PSF at $70\micron$ smears emission from the bright central starburst into the outflow \citep{Bocchio2016}, so measuring \qpah\ in the wind requires careful PSF modeling and removal. JWST $21~\micron$ imaging of the NGC~253 wind could avoid this by providing $R_{\rm PAH} = {\rm (F770W + F1130W) / F2100W}$ at matched resolution, a reliable \qpah\ tracer in nearby galaxies \citep{Chastenet2023_rpah,Egorov2023,Sutter2024}.


\section{Summary \& Conclusions}\label{sec:summary}
We present the first spatially resolved study of PAH evolution in the galactic wind of NGC~253, providing one of the most detailed views of the outflow to date, made possible by the sensitivity and resolution of JWST. Our conclusions are as follows:

\begin{enumerate}

    \item JWST detects PAH (\three, \seven, and \eleven) and continuum emission in both outflow cones, including the limb-brightened streamers along their edges, out to a deprojected distance of at least $r \sim 0.5$~kpc. At $\sim0\farcs05{-}0\farcs375$ ($\sim0.85{-}6.4$~pc) resolution, JWST reveals filamentary sub-structure in the wind not previously resolved (Figures~\ref{fig:pretty_pictures}, \ref{fig:emission_maps}).

    \item The surface brightnesses of \three, \seven, and \eleven\ PAH emission decline faster than $r^{-2}$ in the SW and NE streamers. This steep decline suggests changes in gas column density, DGR, or PAH abundance, rather than geometric dilution from the illumination source alone. Conversely, the NW streamer roughly follows $r^{-2}$, as in M82 \citep{Cronin2026}, suggesting a natural decline in emission with distance from the starburst (Figure~\ref{fig:emission_profiles}).
    
    \item The \ElevenSeven\ PAH band ratio increases with deprojected distance from the nucleus for the SW and NW streamers, consistent with M82, and for the NE before disk subtraction. Unlike M82, \ThreeEleven\ and \ThreeSeven\ decrease for the SW and NE, while increasing for the NW. Comparison with dust models \citep{Draine2021} reveals a general trend toward more neutral PAHs with distance in the SW, NW, and possibly NE streamers, and toward larger PAHs in the SW and NE, with size showing the largest change in the SW. The decreasing ionization may be due to a declining radiation field and ionization parameter, while the shift in size may indicate preferential destruction of small PAHs. The outermost SW and NE ratios match those found in disks and disk-like centers of nearby star-forming galaxies \citep{Pathak2026}, while the innermost ratios match those of star-forming centers and CMZs (Figures~\ref{fig:ratio_profiles}, \ref{fig:draine_models}, \ref{fig:draine_models_no_disksub}).
    
    \item Adopting a deprojected outflow velocity of $\sim 200{-}400$~\kms\ yields $t_{\rm dyn} \sim 1{-}2.5$~Myr for the SW streamer in NGC~253 and $t_{\rm dyn} \sim 5{-}10$~Myr for the inner $\sim 2$~kpc of the M82 wind. The size of PAHs in NGC~253 appears to evolve on faster timescales than in M82. The measured band ratios thus likely probe distinct stages of PAH evolution in winds. The NGC~253 observations may sample an earlier phase when small PAHs are destroyed near the starburst, perhaps by shocks or cosmic rays, whereas the M82 measurements probe a later phase in which the surviving larger PAHs are transported mostly unaffected over several Myr.
    
    \item Assuming a characteristic radius of $r_{\rm cl} \sim 4{-}6.5$~pc and $v_{\rm cool} \sim 200{-}400$~\kms, the cloud-crushing time of the SW streamer is $t_{\rm cc} \sim 0.05{-}0.35$~Myr, meaning it has undergone at least $\sim 2{-}3~t_{\rm cc}$. Radiative cooling is faster ($t_{\rm cool, mix} \sim 10^{-3}~{\rm Myr} \ll t_{\rm cc}$), suggesting cool material in the SW streamer has survived the flow for $\sim 1{-}2.5$~Myr.
\end{enumerate}

\section*{Acknowledgments}
We extend warm thanks to Alyssa Pagan (STScI) for processing the multicolor images in Figure~\ref{fig:pretty_pictures}, Volker Heesen for the $1.5$~GHz image in Figure~\ref{fig:radio}, and Emma Mirizio for helpful discussions.

This work is based on observations made with the NASA/ESA/CSA JWST. The data were obtained from the Mikulski Archive for Space Telescopes at the Space Telescope Science Institute, which is operated by the Association of Universities for Research in Astronomy, Inc., under NASA contract NAS 5-03127 for JWST. All JWST data used in this paper can be found in MAST: \dataset[10.17909/0mv0-e337]{http://dx.doi.org/10.17909/0mv0-e337}. These observations are associated with program JWST-GO-01701. Support for program JWST-GO-01701 is provided by NASA through a grant from the Space Telescope Science Institute, which is operated by the Association of Universities for Research in Astronomy, Inc., under NASA contract NAS 5-03127. This research has made use of NASA's Astrophysics Data System Bibliographic Services. This research used resources of the Oak Ridge Leadership Computing Facility, which is a DOE Office of Science User Facility supported under Contract DE- AC05-00OR22725, using Frontier allocation AST181 and SummitPLUS allocation AST200. 
    
S.A.C., A.D.B., and Y.-H.T. acknowledge support from the NSF under award AST-2108140. V.V. acknowledges support from the Comité ESO Mixto 2024 and from the ANID BASAL project FB210003. H.M.R. acknowledges support from the Flatiron Research Fellowship. The Flatiron Institute is a division of the Simons Foundation. D.P. acknowledges support from the National Science Foundation Graduate Research Fellowship Program under Grant No. DGE-2240614.  H.K, L.H. and K.S. acknowledge funding support from NASA award 80NSSC21K0851, JWST-GO-02987.006-A, and JWST-AR-06061.001-A.
The National Radio Astronomy Observatory and Green Bank Observatory are facilities of the U.S. National
Science Foundation operated under cooperative agreement by Associated Universities, Inc. S.A.C. used Anthropic's Claude Opus 5.5 to optimize and refine code under full human supervision and scrutinization.
\vspace{3mm}
\begin{contribution}
S.A.C. led the project by carrying out the data analysis, producing all figures, and writing the manuscript. A.D.B. was PI of the JWST program and provided guidance for the data reduction and interpretation. R.C.L. led the technical justification and contributed to the observing strategy for the JWST proposal. K.D.G., E.T., and M.L.B. produced the science-ready data products with the JWST pipeline. L.H. and H.B.K. assisted with the continuum subtraction. All coauthors contributed to the manuscript writing, science interpretation, and/or proposal preparation.
\end{contribution}

\facilities{JWST(MIRI, NIRCam)}
\software{\texttt{astropy} \citep{astropy13, astropy18}, \texttt{numpy} \citep{van11, harris2020}, \texttt{matplotlib} \citep{hunter07}, \texttt{scipy} \citep{virtanen20}, \texttt{jupyter} \citep{kluyver16}, JWST Calibration Pipeline \citep{JWSTPipeline_1.12.5,JWSTPipeline_1.17.1}, \texttt{STPSF} \citep{STPSF_github}, \texttt{DOLPHOT} \citep{Dolphin2000, Dolphin2016, Weisz2024}, \texttt{pypher} \citep{Boucaud2016}, \texttt{PAHFIT} \citep{Smith2007}, \texttt{JHAT} \citep{Rest2023}, \texttt{synphot} \citep{synphot}}
\appendix
\vspace{-6mm}
\section{Continuum Subtraction}
\begin{figure*}
    \centering
    \includegraphics[scale=0.5]{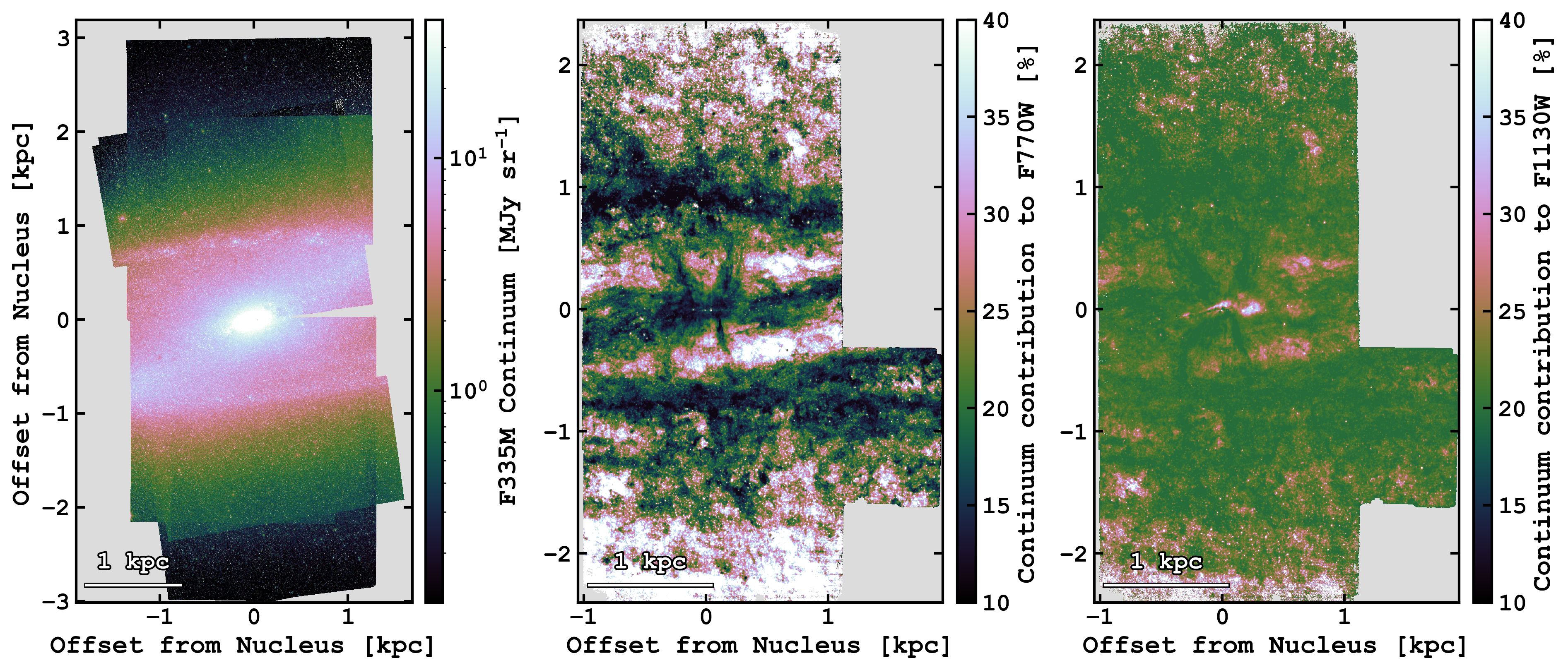}
    \caption{(\textit{Left}) Estimated continuum in the F335M filter as a result of the continuum-subtraction process in Section~\ref{sec:data}. The continuum appears dominated by starlight with no visible outflow emission. (\textit{Middle, Right}) Estimated percentage of continuum contributing to the F770W and F1130W filters after using the approach from \LHands. In the outflow, the median continuum contribution to F770W and F1130W is $\sim 20\%$.}
    \label{fig:continuum}
\end{figure*}

Figure~\ref{fig:continuum} shows the estimated continuum in the F335M, F770W, and F1130W filters. The F335M continuum image is the result of the process described in \citet{Bolatto2024} and \citet{Cronin2026} and summarized in Section~\ref{sec:data}. This process works well, as the resulting continuum image contains many point sources and no residual wind emission. We also tested this method against the empirical prescription derived in \citet{Koziol2026}, which isolated PAH-correlated emission (determined from F1130W) in the F335M filter using the bracketing F300M and F360M filters. Substituting our F250M data in for F300M, and tweaking the F1130W intensity threshold that determines PAH-correlated emission, we produced a continuum-subtracted F335M image that is consistent with results when assuming a lower $R$ that marks where the F335M continuum lies between the flanking filters ($R \approx 0.6$) in the \citet{Bolatto2024} method. To preserve contiguous streamers, we prefer the \citet{Bolatto2024} method with $R = 0.75$. 

Also in Figure~\ref{fig:continuum}, we display the percentage of continuum in the F770W and F1130W filters, estimated using the method in \LHands, which we summarize in Section~\ref{sec:data}. In the wind, the continuum makes up $\sim 20\%$ of the total emission measured in F770W and F1130W.

\begin{figure*}
    \centering
    \includegraphics[scale=0.35]{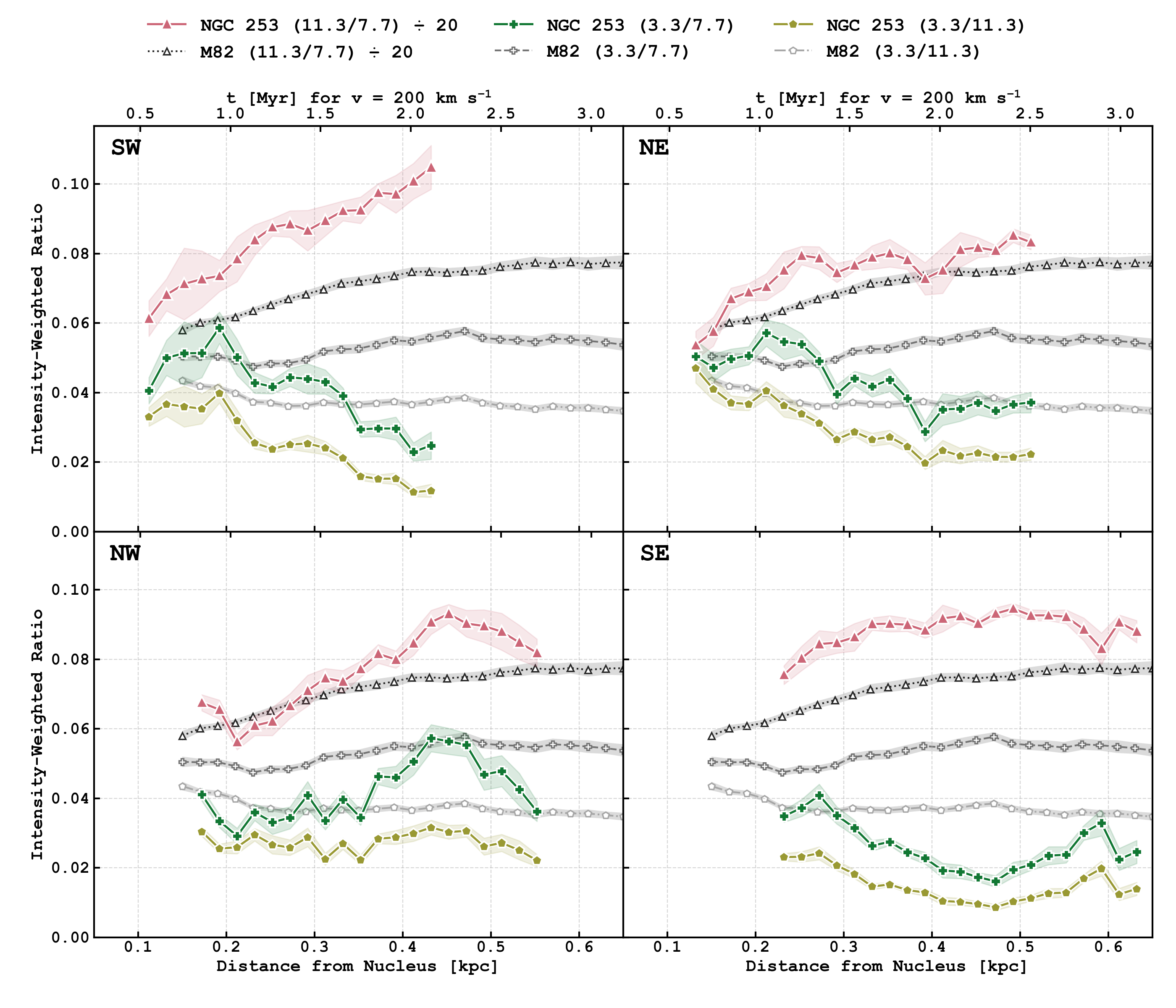}
    \caption{The same as Figure~\ref{fig:ratio_profiles}, except no disk subtraction is applied to the PAH emission measured in the wind, and we include the SE streamer. Removing the estimated disk emission injects some fluctuations into the NE profiles, but the overall trends in the SW, NE, and NW streamers match Figure~\ref{fig:ratio_profiles}. The SE streamer shows a slight increase in \ElevenSeven\ and a decrease in the \three\ ratios before reaching the spiral arms around $r \sim 0.6$~kpc.}
    \label{fig:ratio_profiles_no_disksub}
\end{figure*}

\begin{figure*}
    \centering
    \includegraphics[scale=0.4]{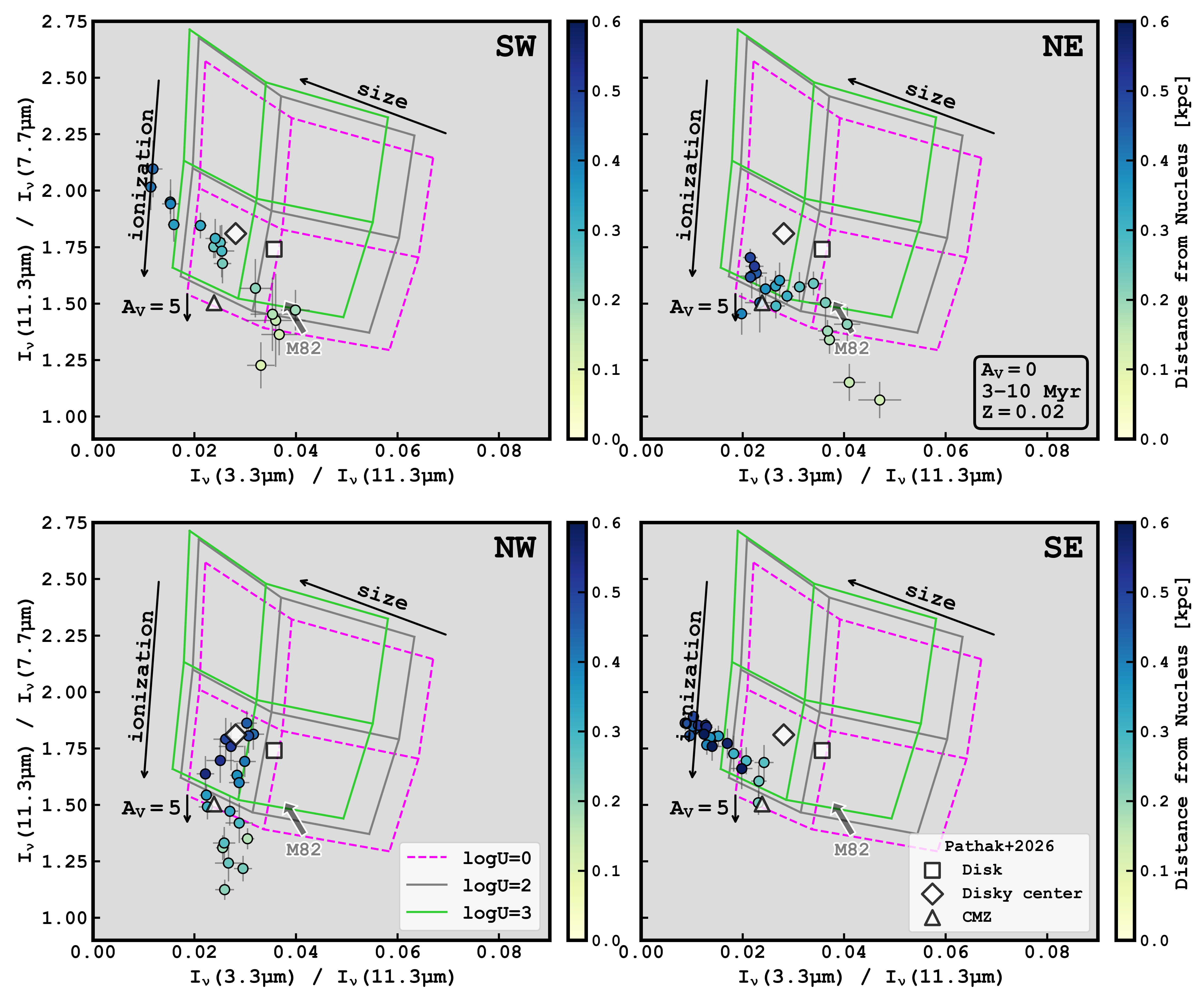}
    \caption{The same as Figure~\ref{fig:draine_models}, except no disk subtraction is applied to the PAH emission measured in the wind, and we include the SE streamer. The overall trends in the SW, NE, and NW streamers match Figure~\ref{fig:draine_models}. Removing the estimated disk emission shifts \ThreeEleven\ to larger values (i.e., toward smaller PAHs) by up to ${\sim}0.2$~dex, while \ElevenSeven\ changes by $\lesssim 0.03$~dex. The disk+wind emission along the SE streamer matches the trends observed in the SW and NE.}
    \label{fig:draine_models_no_disksub}
\end{figure*}

\section{Non-Disk-Subtracted PAH Band Ratios}
As discussed in Section~\ref{sec:results}, the inclination of the NGC~253 disk results in each line of sight containing a mixture of wind and disk emission. We correct for this by estimating pure disk emission in slices taken adjacent to the streamers (see Figure~\ref{fig:emission_maps}) to produce the disk-corrected wind emission profiles in Figure~\ref{fig:emission_profiles} and the disk-corrected PAH ratios in Figures~\ref{fig:ratio_profiles} and \ref{fig:draine_models}. Here we show the uncorrected (wind+disk) ratios. Figure~\ref{fig:ratio_profiles_no_disksub} displays the PAH ratio profiles, and Figure~\ref{fig:draine_models_no_disksub} marks where these ratios fall on the theoretical dust model grids of \citet{Draine2021}.

Figure~\ref{fig:ratio_profiles_no_disksub} reveals a general trend toward higher \ElevenSeven\ with distance from the starburst in the SW, NE, and NW streamers, and a decreasing \ThreeSeven\ and \ThreeEleven\ in the SW and NE streamers. SE exhibits similar behavior, with turnover in all ratios at further radii likely due to confusion with the spiral arms, and the NW streamer shows a flatter \ThreeSeven\ and \ThreeEleven. Comparison to the \citet{Draine2021} models reveals larger, more neutral PAHs with distance in the SW, NE, and SE streamers; the size distribution does not change in the NW streamer. Though some scatter is introduced when calculating the disk-subtracted, outflow-only ratios, the uncorrected ratio trends are consistent with the results in Figures~\ref{fig:ratio_profiles} and \ref{fig:draine_models}. This agreement suggests that the trends in the uncorrected ratios are driven primarily by the wind rather than the disk, lending some confidence in the trends seen in the SE streamer, for which we were unable to provide a reliable disk subtraction.

\begin{figure*}
    \centering
    \includegraphics[scale=0.5]{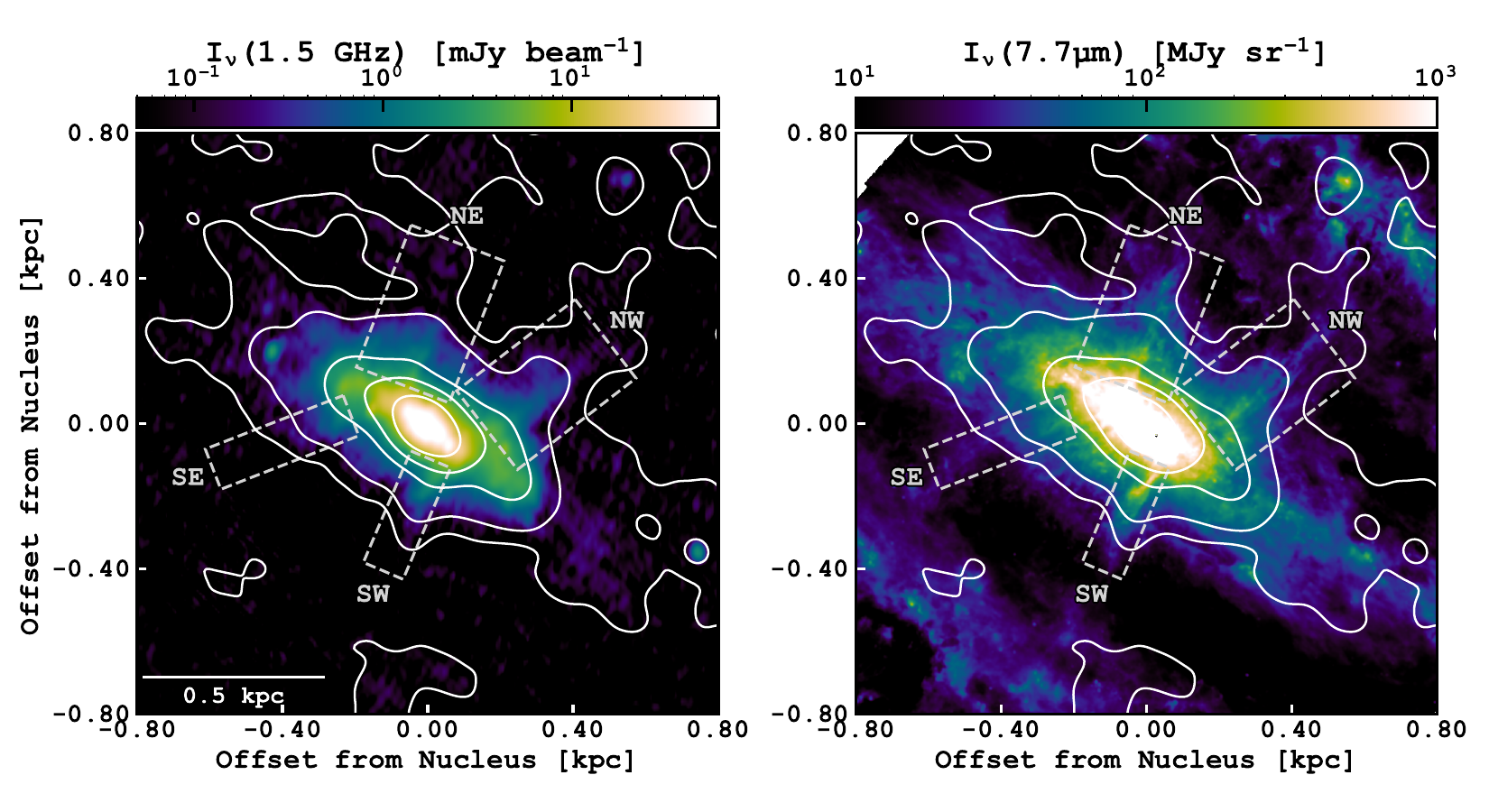}
    \caption{Synchrotron emission overlaps with PAHs in the NGC 253 wind. (\textit{Left}) $1.5$~GHz VLA map of NGC 253 from \citet{Heesen2011}, with contour levels at 0.03, 0.2, 1.8, 6, and 20 mJy~beam$^{-1}$ with Gaussian smoothing to $\sim 5''$, and boxes representing the outflow streamer locations. (\textit{Right}) \seven\ PAH map with the same contours and streamer locations. The streamers show similar morphologies and brightness gradients in synchrotron and PAH emission, perhaps indicating that the cosmic rays and shocks traced by the synchrotron emission influence the PAH emission.}
    \label{fig:radio}
\end{figure*}

\section{Co-Spatiality between Synchrotron Emission and PAHs}
Preferential destruction of smaller PAHs may be due to processes such as shocks or cosmic rays. Theoretical predictions show that smaller PAHs are destroyed faster than larger PAHs with shocks $\gtrsim 100$~\kms\ \citep{Micelotta2010shocks}, and cosmic ray ions and electrons can penetrate dense clouds and destroy small PAHs ($N_{\rm C} = 50$) on timescales $\gtrsim 20$~Myr for solar-neighborhood cosmic ray intensities \citep{Micelotta2011}. In the central starburst of NGC~253, the cosmic-ray ionization rate is $\sim 10^2{-}10^3\times$ higher than in the solar neighborhood \citep{Behrens2024}, making this timescale $\lesssim 1$~Myr. Moreover, cosmic rays have been found to be advected out by the outflow \citep{Heesen2011, Wang2026}. In Figure~\ref{fig:radio}, we show the co-spatiality between $1.5$~GHz synchrotron emission from \citet{Heesen2011}, tracing cosmic-ray electrons and shock-compressed magnetic fields, and \seven\ PAH emission. We overlay both maps with $1.5$~GHz contour levels at 0.03, 0.2, 1.8, 6, and 20 mJy~beam$^{-1}$ from the VLA map Gaussian-smoothed to $\sim 5\arcsec$. Dashed rectangles represent the locations where we estimate streamer emission. We find good agreement in both morphology and brightness gradients between the synchrotron emission and PAHs, hinting at possible interactions between cosmic rays, shocks, and PAHs.

\bibliography{bib}{}
\bibliographystyle{aasjournalv7}


\end{document}